# Addictive Auctions: using lucky-draw and gambling addiction to increase participation during auctioning

Ravin Kumar[0000-0002-3416-2679]

Department of Computer Science, Meerut Institute of Engineering and Technology, meerut-250005, Uttar Pradesh, India
`ravin.kumar.cs.2013@miet.ac.in`

**Abstract.** Auction theories are believed to provide a better selling opportunity for the resources to be allocated. Various organizations have taken measures to increase trust among participants towards their auction system, but trust alone cannot ensure a high level of participation. We propose a new type of auction system which takes advantage of lucky draw and gambling addictions to increase the engagement level of candidates in an auction. Our system makes use of security features present in existing auction systems for ensuring fairness and maintaining trust among participants.

## 1    Introduction

Auction systems are used for allowing authorities to fairly allocate resources among participants based on some defined set of rules. It is a common belief that auction system helps in allowing seller to gain more profit as it allows groups to compete with each other to obtain a certain resource. Different types of auction systems are developed to maintain the trust of participants in auction systems with the hope of obtaining better selling price for the resource. While in other systems such as gambling, a large number of people lose money but still most of the people got addicted to it and keep playing. Similarly, in a lottery-based system, people know that their chance of winning is the same as everyone else but they keep believing that they might win and continue buying lottery tickets. Despite the fact that in gambling and lottery-based systems the majority of the people are destined to losses money without getting any reward, still the participation number is very high with respect to auctions where if someone is allocated the resource than only, they have to pay their bidding amount

In this paper, we explored more on the integration of an auction system with gambling, and lucky-draw based addictions to design a hybrid type of auction system. This hybrid system contains addictive nature present in lucky draw and gambling systems. We have also studied the effects of our proposed auction systems on the engagement of participants and on the selling price of the auctioned resources

## 2    Related Works

Y.H. Lee et al. [1] studied the effect of prospect imageability and mental imagery on consumer behavior. H Wardle et al. [2] analyzed the effect of gambling machines on social and economic condition of its locality. A Blaszczynski et al. [3] described general methods which can be used to reduce the negative impact of gambling and focuses on responsible form of gambling. V Ariyabuddhiphongs et al. [4] explained various forms of gambling and focused on reason which motivates people to buy lottery and against a popular belief he further concluded that big wins tends to improve winner's life. Y Mu et al. [5] proposed an internet-based bidding system for sealed-bid auctions fulfilling necessary security requirements without using multiple server systems. CY Olivola et al. [6] used incentive compatible mechanism to compare time vs money preference behavior. WJ Jung et al. [7] focused on studying customer engagement using contest based social media marketing. R Zhang et al. [8] proposed a quantum system ensuring security in sealed-bid auction. G E Kersten et al. [9] performed a comparative study on auctions and negotiations and then suggested that social effect become much dominant over competitive nature when transparency is provided. AK Naik et al. [10] suggested that auctioneer also plays important role in bidding and some benefits should be given to auctioneers to keep their motivation high which will lead to higher final bidding prices.

## 3    Proposed auction system

Addictive auction system consists of two fundamental concepts of catalysts and recipients. The catalysts are the participants who provides some amount of money to the recipients (Fig. 1).

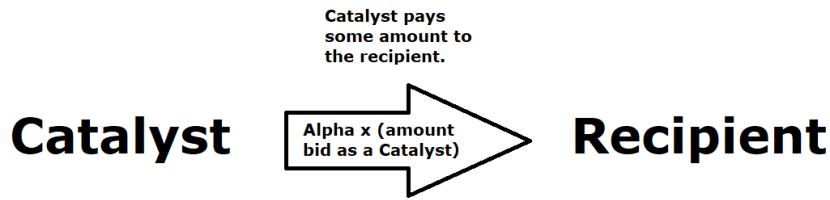

**Fig. 1.** Catalyst pays some amount to the recipient.

In our method, there is a dynamic list containing $P_{id}$ and its association with candidates and their bidding amount. Each $P_{id}$ contains details about the amount bid at an instance, along with the bidder details. This dynamic list keeps on updating with every new bidding during the auctioning. Each $P_{id}$ gets updated with every new bidding amount introduced by any participant during the auctioning. The $P_{id}$ of each individual is updated using a simple formula.

$$P_{id} = T - n_i$$

Here, T represents the total number of biddings that took place up to that instant, and $n_i$ represents the instances at which each person had bid in the auction. To better understand the concept of $P_{id}$ , we have provided a sample demonstration.

**Table 1.** Sample demonstration using Catalyst at id = 3 and recipient at id = 0

| Instance | $P_{id}$ and associated details | Catalyst | Recipient |
|---|---|---|---|
| 0 | P0: {Person 1, 100} | Null | Person 1 |
| 1 | P0: {Person 2, 150}, P1: {Person 1, 100} | Null | Person 2 |
| 2 | P0: {Person 3, 200}, P1: {Person 2, 150}, P2: {Person 1, 100} | Person 1 | Person 3 |
| 3 | P0: {Person 1, 250}, P1: {Person 3, 200}, P2: {Person 2, 150}, P3: {Person 1, 100} | Person 1 | Person 1 |
| 4 | P0: {Person 3: 400} P1: {Person 1, 250}, P2: {Person 3, 200}, P3: {Person 2, 150}, P4: {Person 1, 100} | Person 2 | Person 3 |

Our system also has a parameter α with a range of (0,1]. This parameter plays a major role in deciding the amount to be paid to the recipient. Before the beginning of the auction, values of α and index of Catalyst (i.e. Pc) are already known to all the participants. Catalyst and recipients are assigned such that the id of catalyst is always greater than that of the recipient. The person with $P_{id}$ of a catalyst will provide alpha times the amount associated with its id to the recipient. Each new bidding increases the amount that participant with catalyst id have to pay to the recipient.

Since the probability of one becoming a catalyst in total of n+1 participant is 1/(n+1) which is much lower than the possibility of one losing a one-winner based gambling i.e. n/(n+1). As the auction progresses, our system motivates more people to take advantage of becoming a recipient. While with each increasing bidding it motivates participants with the catalyst $P_{id}$ to bid higher and move away from the catalyst position.

We have provided two varieties of addictive auction systems based on the relationships present among the catalysts and recipients and source code for both these variants are available at our github repository [11].

### 3.1 When recipient is the highest bidder

In this type of addictive auction, the recipient is the highest bidder. This type of system motivates the participants to make the highest bidding and take advantage of receiving some amount from the catalyst as well as obtain the auctioned resource (Fig. 2).

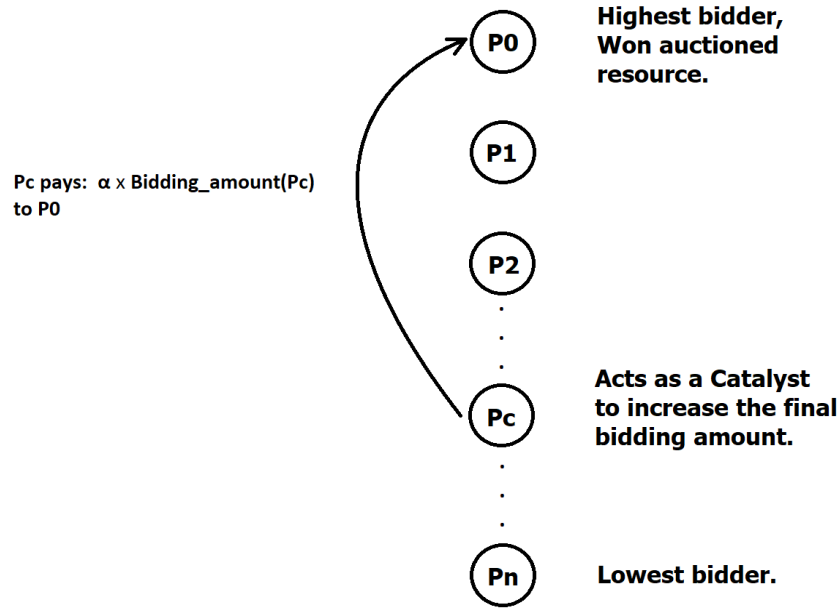

**Fig. 2.** When highest bidder is assigned as the recipient.

### 3.2 When recipient is in between the highest bidder and the catalyst

In this arrangement the highest bidder is not the recipient, which makes this arrangement a pure hybrid of a traditional auction system and lucky-draw based gambling system (Fig. 3). This type of arrangement motivates the candidates to participate in the auction. This motivation leads to an increase in the frequency of bidding, and in turn helps in increasing the final bidding amount.

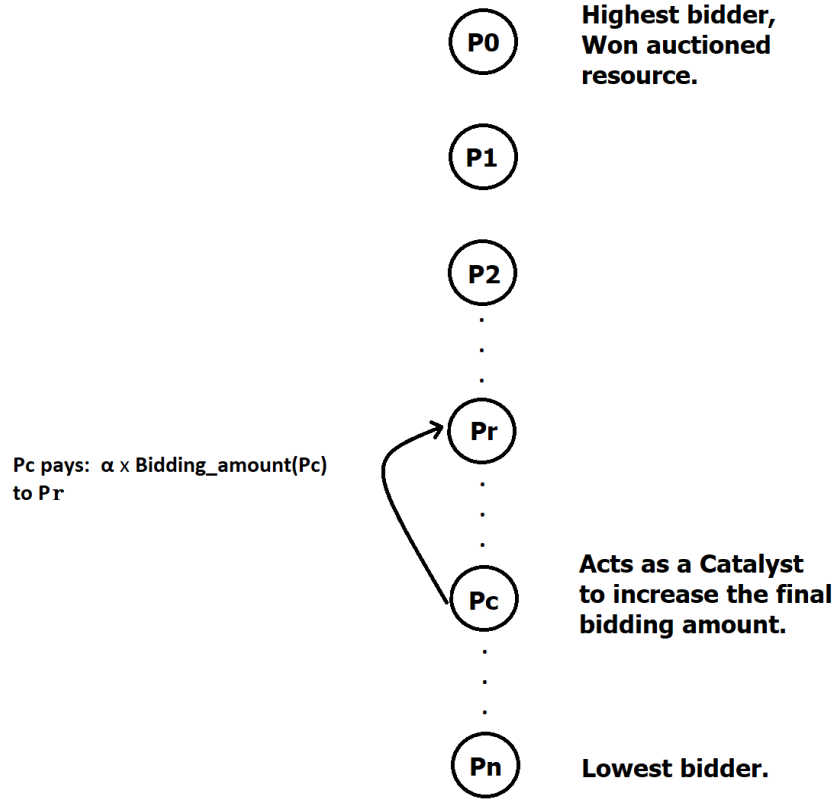

**Fig. 3.** When the recipient is in between the highest bidder and the catalyst.

Due to this hybrid nature of our proposed system, it can also be named as "collaborative auction" as the addictive nature of gambling helps the candidates to participate at a much higher level.

## 4       Working Demonstration

We conducted an auctioning experiment to analyze the effects of both variants of our proposed auctioning system. In this experiment we studied the differences in the final bidding amount, engagement of participants, and in the total time taken to reach the final bidding among general auctioning and our addictive auctioning system.

**Table 1.** Experiment related details

| S.no | Symbol/Variable name | Description |
| --- | --- | --- |
| 1 | Total Participants | 50 |
| 2 | Parameter α | 0.1 |
| 3 | Catalyst id (in 4.2) | 4 |
| 4 | Catalyst id (in 4.3) | 4 |
| 5 | Recipient id (in 4.3) | 2 |

### 4.1 When a highest-bidder wins, type traditional auction system is used

In this case, auctioning is performed using traditional rule of highest-bidder gets the auctioned resource. After repeated experimentation with different groups of participants an average bidding trend in obtained (Fig. 4)

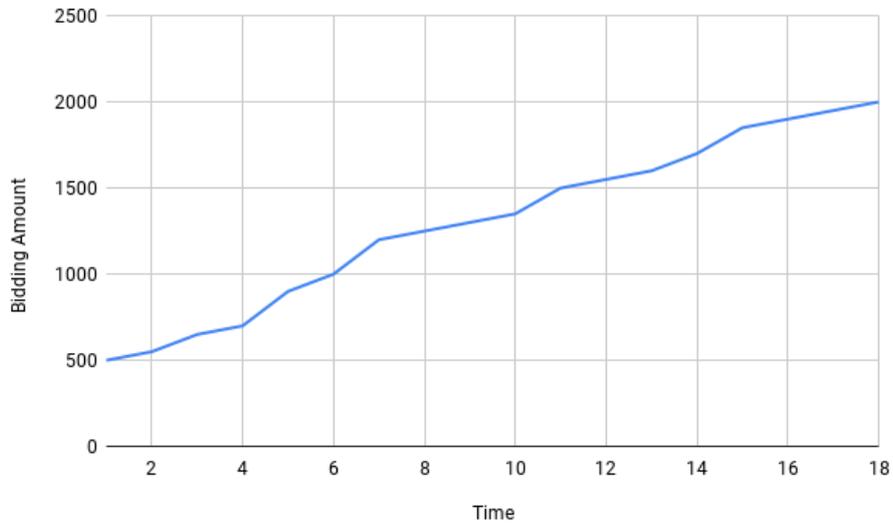

**Fig. 4.** Visual analysis of bidding data.

### 4.2 Addictive auction, when recipient is the highest bidder

In this case, bidding is performed using the recipient is the highest bidder variant of proposed addictive auction system. The average bidding trends obtained in this variant shows much higher bidding amounts, suggesting that the system motivated candidates to bid higher and higher as the highest bidder will also be having benefit of being the recipient (Fig. 5).

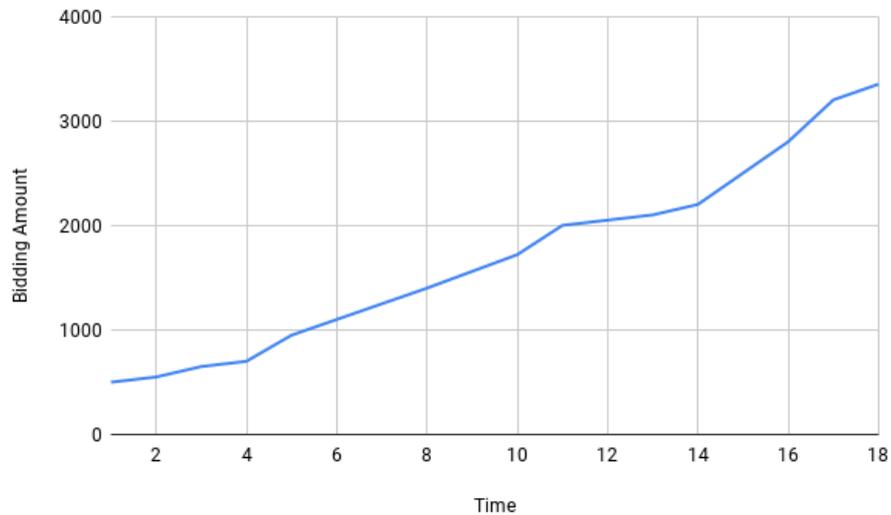

**Fig. 5.** Visual analysis of bidding data.

### 4.3 Addictive auction, recipient is between highest bidder and catalyst

In this case, bidding is performed using the recipient in between highest bidder and catalyst variant of proposed addictive auction system. After analyzing the average bidding trends, It suggests that although the system motivated candidates to bid higher but after some point the rapid increase in the bidding price slows down a little, as some participants became interested in obtaining recipient Pid and taking free money at home (Fig. 6).

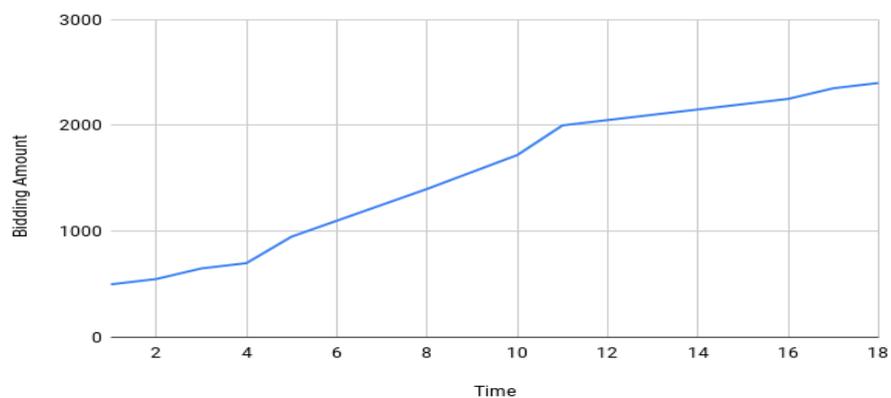

**Fig. 6.** Visual analysis of bidding data.

### 4.4    Overall Comparative analysis

A comparative analysis on the collected auction data is performed to figure out the performances of traditional highest-bidder wins and variants of addictive auction systems

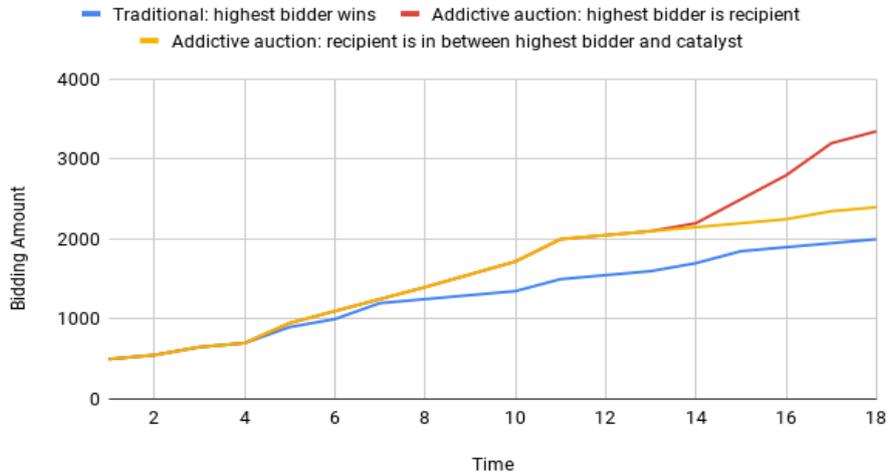

**Fig. 7.** Visual analysis of bidding data.

After analysis (Fig. 7), it became clear that both variants of proposed addictive auctions took less time to achieve the final bidding amount of traditional highest-bidder wins. This happened because of the addictive nature of gambling (i.e. catalyst and recipient) kept candidates more engaged and motivated to take part in the auction.

Although, among both the variants of addictive auctions when highest-bidder was kept the recipient had reached higher final bidding amount than the other. This might have happened because in the longer run, some participants became more interested in becoming the recipient and not the highest-bidder. This situation could be further resolved by kept on increasing the minimum increment to amount for doing bidding as the auction goes forward.

## 5    Conclusion

Our proposed auction systems have shown potential of increasing the overall participation in auctions leading to increase in the final bidding amount. The combined effect of addictive nature of gambling system to an auction system help in increasing the participation of candidates. It also makes auctions more attention gathering and in turn reducing the illegal gambling activities which in turn leads to better law and order establishment.